%% file: nonlinccfm.tex
\documentclass{article}
\input{common_preamble.tex}

\title{
\begin{flushright}
{\small IFJPAN-IV-2013-11}  \\
\end{flushright} $ $ \\
Gluon saturation scale from the KGBJS equation}
\author{Krzysztof Kutak, Dawid Toton
\\ \vspace*{0.5cm} \\
{\it Instytut Fizyki Jądrowej im. H. Niewodniczańskiego} \\
{\it Radzikowskiego 152, 31-342 Kraków, Poland}
}

\begin{document}
\maketitle

\begin{abstract}
The CCFM equation and its extended form with a quadratic term
(KGBJS equation)
are solved with fixed and running coupling constant.
The solution of the KGBJS equation is compared to gluon densities resulting from the CCFM and BK equations.
As the saturation scale $Q_s$ now becomes available as a function of the hard scale $p$
 we observe that low values of $p$ impede its growth with $\frac 1 x$.
Also, at values much larger than partons transversal momentum
the saturation effects become independent on the hard scale what we call liberation of saturation scale.
We also introduce the hard-scale-related saturation scale $P_s$ and investigate its energy dependence.
We observe that the new scale as a function of $x$
decreases starting from the value of transversal momentum of gluon.
\end{abstract}
\section{Introduction}
We consider hadronic scattering in the limit of high center-of-mass energy, where the energy is the largest scale in the problem.
Perturbative treatment of processes with high momentum transfer at high energies leads to decomposition of the cross section into hard matrix element and gluon density \cite{Gribov:1984tu,Catani:1990eg} which is a function of the longitudinal momentum fraction $x$ and transverse momentum ${\bf k}$ of a gluon as well as a scale $p$ related to a hard process.
The gluon density obtained in such a setup at not too large parton densities obeys the CCFM \cite{Ciafaloni:1987ur,Catani:1989sg,Catani:1989yc} equation. The equation sums up gluons with a condition of strong ordering in angle and can be viewed as a bridge between BFKL and DGLAP regimes.
The particularly interesting is however to apply the CCFM framework to saturation physics \cite{Gribov:1984tu} in order to investigate saturation with the help of exclusive processes like for example di-jet production at the LHC \cite{Deak:2009xt,Deak:2009ae,Deak:2010gk,Kutak:2012rf,Chatrchyan:2012gwa}.
The first step towards introducing saturation in the CCFM framework has been done in \cite{Kutak:2008ed,Avsar:2009pv,Avsar:2009pf} applying the absorptive boundary method  \cite{Mueller:2002zm} in order to suppress gluon density at low values of gluon's transversal momentum ${\bf k}$. Another approach has been developed in \cite{Kutak:2011fu,Kutak:2012yr,Kutak:2012qk} where the extension of CCFM to allow for dynamical gluon saturation has been proposed. The proposed equations (for Weizsäcker-Williams gluon density (KGBJS) in \cite{Kutak:2011fu,Kutak:2012yr} and unintegrated gluon density in \cite{Kutak:2012qk}) have structure similar to the resummed BK equation \cite{Kutak:2011fu} since the form factors in the new equations (Sudakov and non-Sudakov) are linked by the limit procedure to the Regge form factor being present in the resummed BK equation.

In this paper we solve the KGBJS and CCFM equations with running and fixed coupling constants in order to perform realistic phenomenology applications in the future.
We also study in detail the nonlinear effects by calculating the emergent saturation scale in the KGBJS equation. The novel feature of the saturation scale is its nontrivial dependence on the hard scale related variable $p$. We observe that when the
hard scale related variable is much larger than the ${\bf k}$ of gluon the saturation scale stops to depend on it and the BK limit is reached. We call this effect liberation of saturation scale due to relaxing of phase space constraint.
The paper is organized as follows. In the section two we present solutions of the KGBJS and CCFM equations in case of fixed and running coupling constants and study the effect of running coupling on the solutions. In section three we study the saturation effects in the KGBJS equation by analyzing the properties of the saturation scale as emerged due to nonlinearities. We compare that scale to the one generated via the BK evolution \cite{Balitsky:1995ub,Kovchegov:1999yj}. As the equation depends on a hard scale we also introduce hard scale related saturation scale $P_s$.
\section{The CCFM evolution equation and its nonlinear extension}
\subsection{Hard emissions approximation and running coupling effects }
The KGBJS equation reads\footnote{In the nonlinear term we did not include the $1/(1-z)^2$ as it has been suggested in \cite{Deak:2012mx}. In the present paper we are going to solve the equation in the approximate form where the eventual problem observed in \cite{Deak:2012mx} does not show up. We are going to address the problem of proposed modification of the solution of the full equation in the future.}:
\begin{align}
\label{eq:final1}
\mathcal{E}(x, k^2, p)&=\mathcal{E}_0(x,k^2,p)\\\nonumber
&
+\int \frac{d^2\bar{\bf{q}}}{\pi \bar{q}^2}\int_{x/x_0}^{1-Q_0/\overline q} dz
 \,\theta (p - z\bar{q})P_{gg}(z,k,\bar q, p)
\Bigg[
\mathcal{E}\left(\frac{x}{z}, k^{'2}, \bar{q}\right)\\\nonumber
&
-\frac{1}{\pi R^2}\bar{q}^2\delta(\bar{q}^2-k^2)\,\mathcal{E}^2(\frac{x}{z},\bar{q}^2,\bar{q})\Bigg]
\end{align}
where
\begin{equation}
P_{gg}(z,k,\bar q, p)=\bar\alpha_s\Delta_s(p,z\bar{q})
\left ( \frac{\Delta_{ns}(z,k, \bar q)}{z} + \frac{1}{1-z} \right ).
\end{equation}
\begin{figure}[t!]
\centerline{
  \setlength{\unitlength}{1mm}
  \begin{picture}(65,65)
    \put(0,0){\includegraphics[width=6.5cm]{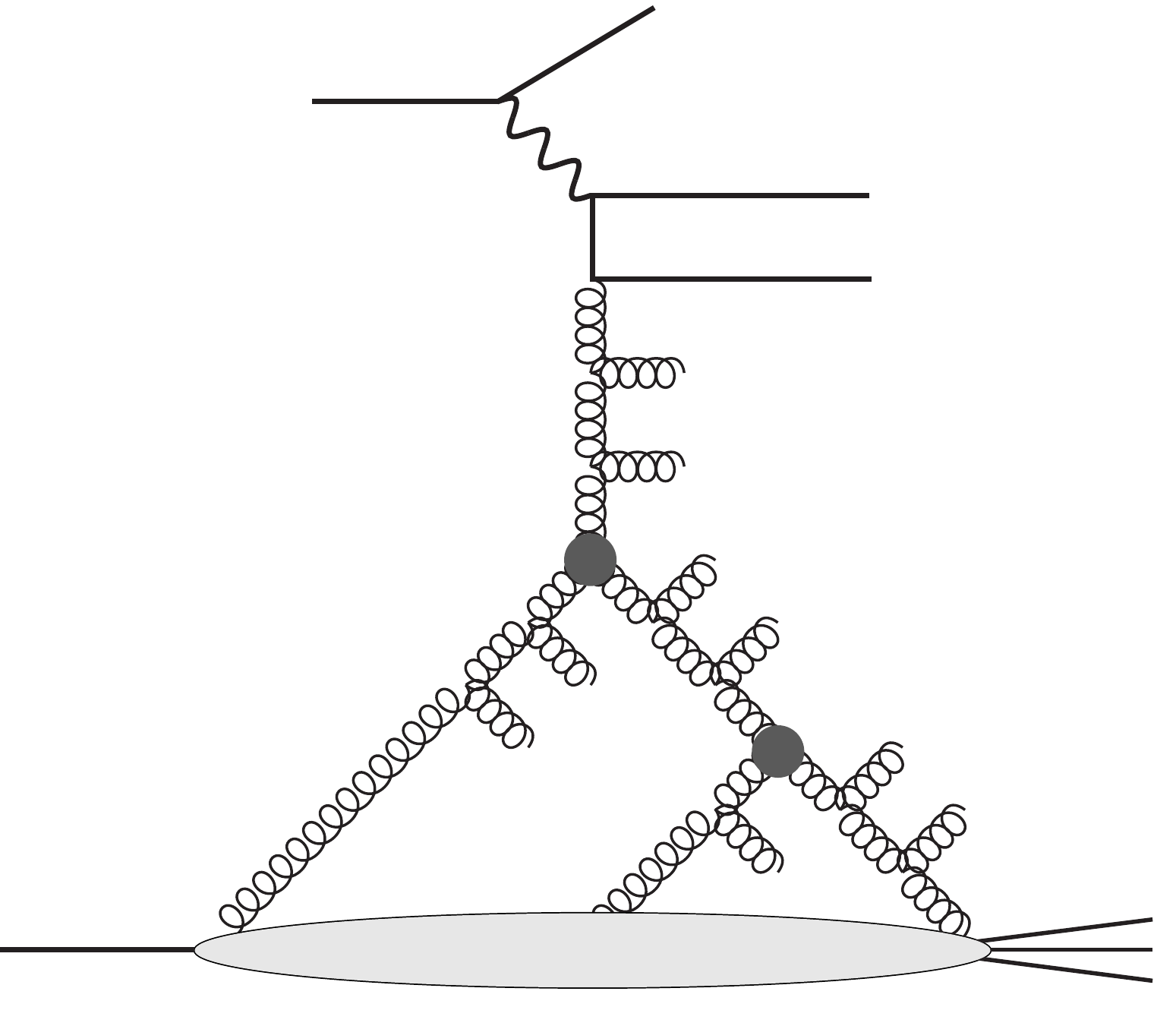}}
    \put(18,48){\makebox(0,0)[lb]{\smash{$p_e$}}}
    \put(25,45){\makebox(0,0)[lb]{\smash{$Q^2$}}}
    \put(24,37){\makebox(0,0)[lb]{\smash{$x$, $k$}}}
    \put(23.5,32){\makebox(0,0)[lb]{\smash{$\frac x z$, $k'$}}}
    \put(39,35){\makebox(0,0)[lb]{\smash{q}}}
    \put(1,1.5){\makebox(0,0)[lb]{\smash{$p_P$}}}
  \end{picture}
}
\caption{Schematic illustration of kinematical variables used in the Eq. \ref{eq:final1}}
\label{diagram}
\end{figure}
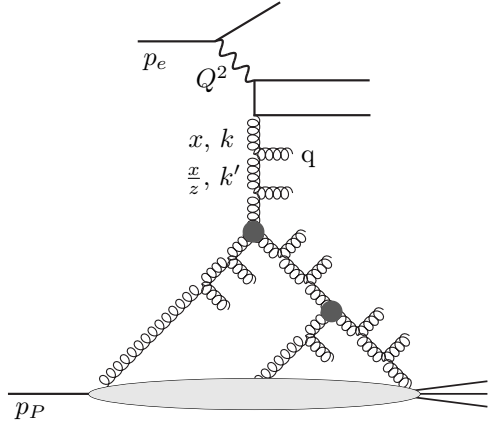
The momentum vector associated with $i$-th emitted gluon is
\begin{equation}
q_i=\alpha_i\,p_P+\beta_i\,p_e+q_{t\,i}.
\end{equation}
The variable $p$ in (\ref{eq:final1}) is defined via $\bar{\xi} = p^2/(x^2s)$ where $\frac{1}{2}\ln(\bar{\xi})$ is a maximal rapidity which is determined by the kinematics of hard scattering,
$\sqrt{s}$ is the total energy of the collision and $k' = |\pmb{k} + (1-z)\bar{\pmb{q}}|$, $\bar\alpha=N_c\alpha_s/\pi$. We also define $k\equiv |{\bf k}|$.
The momentum ${\bf\bar{q}}$ is the transverse rescaled momentum of the emitted gluon, and is related
to ${\bf q}$ by $\bar{{\bf q}} = {\bf q}/(1-z)$ and $\bar q\equiv|{\bar{\bf q}}|$,
and $Q_0$ is a cutoff on gluons momentum.\\
The form factor $\Delta_s$ accompanies the $1-z$ pole and it reads:
\begin{equation}
\Delta_s(p,z\bar{q})=\exp\left(-{\bar\alpha_s}\ln\frac{p}{z \overline q}\ln\frac{z \overline q p}{Q_0^2}\right)
\end{equation}
while the form factor $\Delta_{ns}$ accompanying the $1/z$ pole accounts for angular ordering. We use its form as proposed in \cite{Kwiecinski:1995pu}:
\begin{equation}
\Delta_{ns}(z, k, q) =
\exp\left(
- \bar\alpha_s
\ln\frac{z_0}z
\ln\frac{k^2}{z_0 z q^2}
\right)
\end{equation}
where $z_0 = \frac k q$ for $z < \frac k q < 1$
and outside the interval it assumes the bounding values,
$z_0 = z$ when $\frac k q < z$ and $z_0 = 1$ when $\frac k q > 1$.
The more inclusive form of the equation above valid in the low $x$ regime follows
if we set the Sudakov form-factor $\Delta_s$ to unity
and neglect the contribution from the soft emissions
i.e. $\frac 1 {1-z}$ pole in $P_{gg}$ (and no $z$ cutoff).
We obtain:
\begin{align}
\label{eq:kgbjs2}
\mathcal{E}(x, k, p) &=
\mathcal{E}_0(x, k, p) \\\nonumber
&+
\int_x^{x_0} \frac{\dd w} w
\int_0^\infty \frac {\dd q^2}{q^2}
\int_0^\pi \frac{\dd \phi}{\pi}
\theta\left(p - z q\right)
P_{gg}(z, k, q)
\mathcal{E}\left(w, k', q\right)
\\\nonumber
&-
\frac{1}{\pi R^2}
\int_x^{x_0} \frac{\dd w} w
\theta(p - z k)
P_{gg}(z, k, k)
\mathcal{E}^2\left(w, k, k\right)
\end{align}
where $z = \frac x w$ under both of the $\dd w$ integrals (from now on when we will use the KGBJS acronym we will refer to equation \ref{eq:kgbjs2}).
The splitting function, with running $\alpha_s$ following \cite{Kwiecinski:1995pu}, is simplified to:
\begin{equation}
P_{gg}(z, k, q) =
\bar\alpha_s(k^2)
\frac {\Delta_{ns}(z, k, q)}z.
\end{equation}
The parameter characterizing the target is chosen to be $R=10/\pi$ and the starting point of evolution is chosen to be $x_0=10^{-2}$.
\begin{figure}[t!]
\centerline{
 \includegraphics[width=6.5cm,trim=0.75cm 0 4cm 0]
 {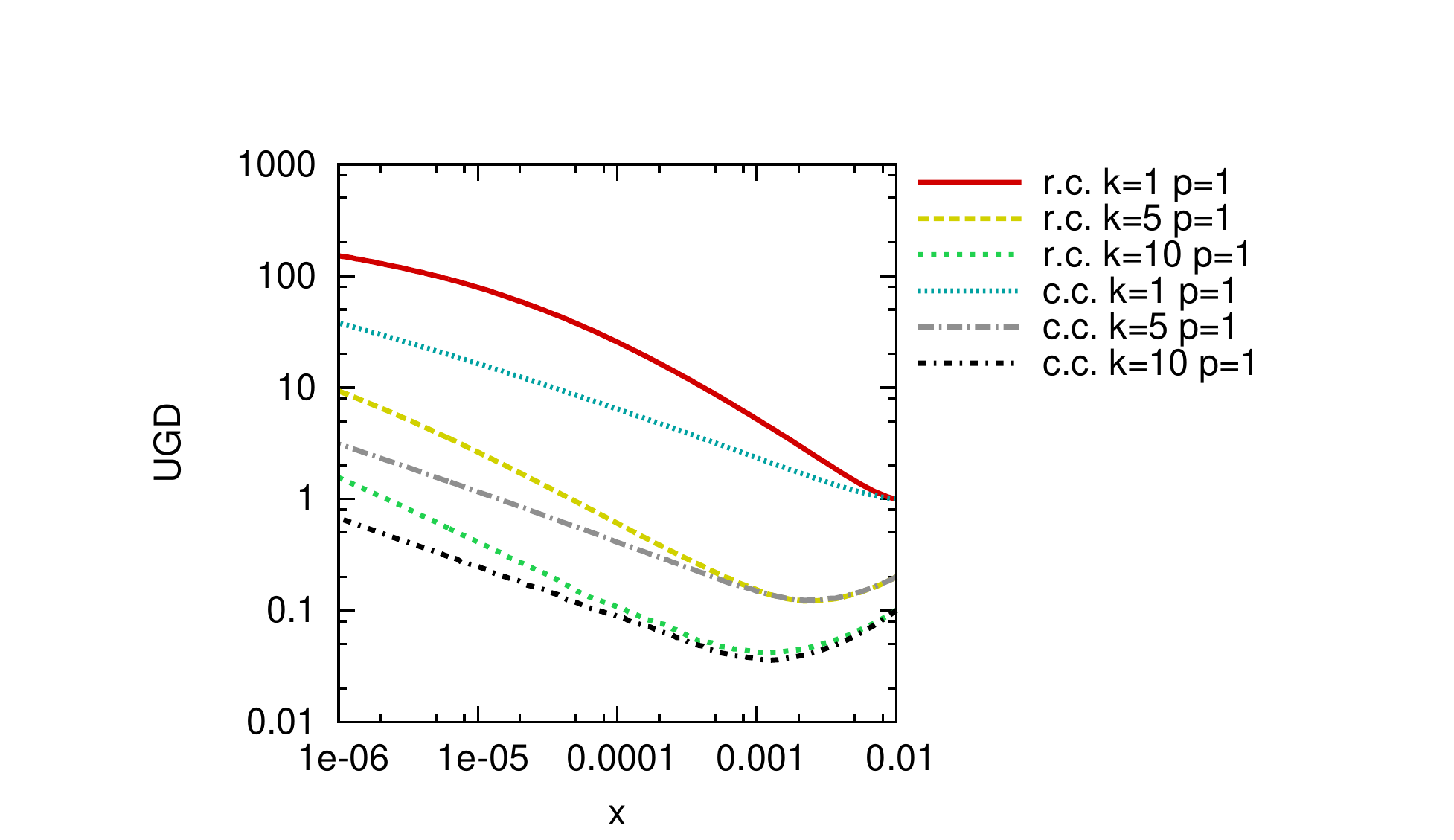}
 \includegraphics[width=6.5cm,trim=0.75cm 0 4cm 0]
 {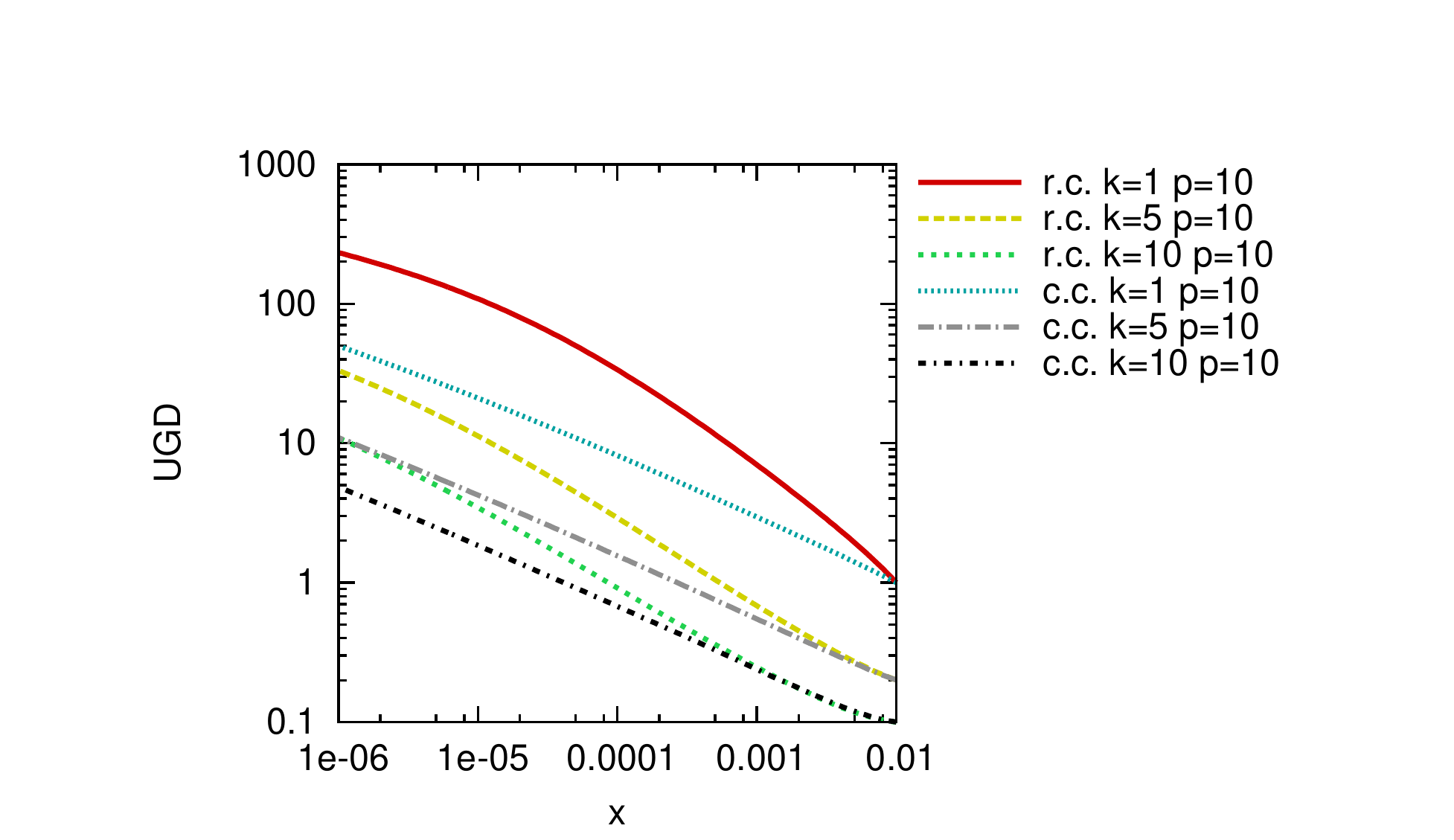}
}
\centerline{
 \includegraphics[width=6.5cm,trim=0.75cm 0 4cm 0]
 {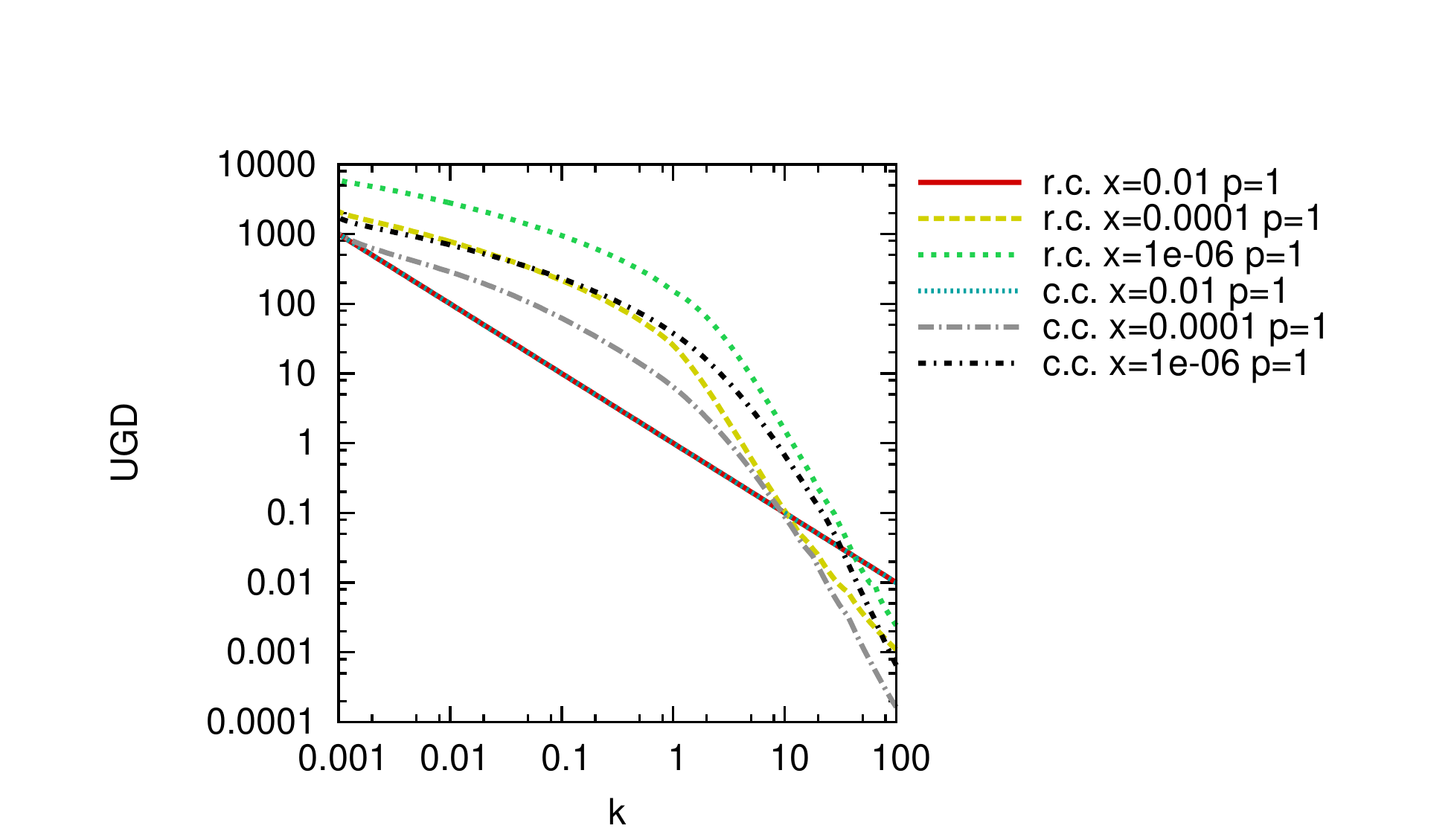}
 \includegraphics[width=6.5cm,trim=0.75cm 0 4cm 0]
 {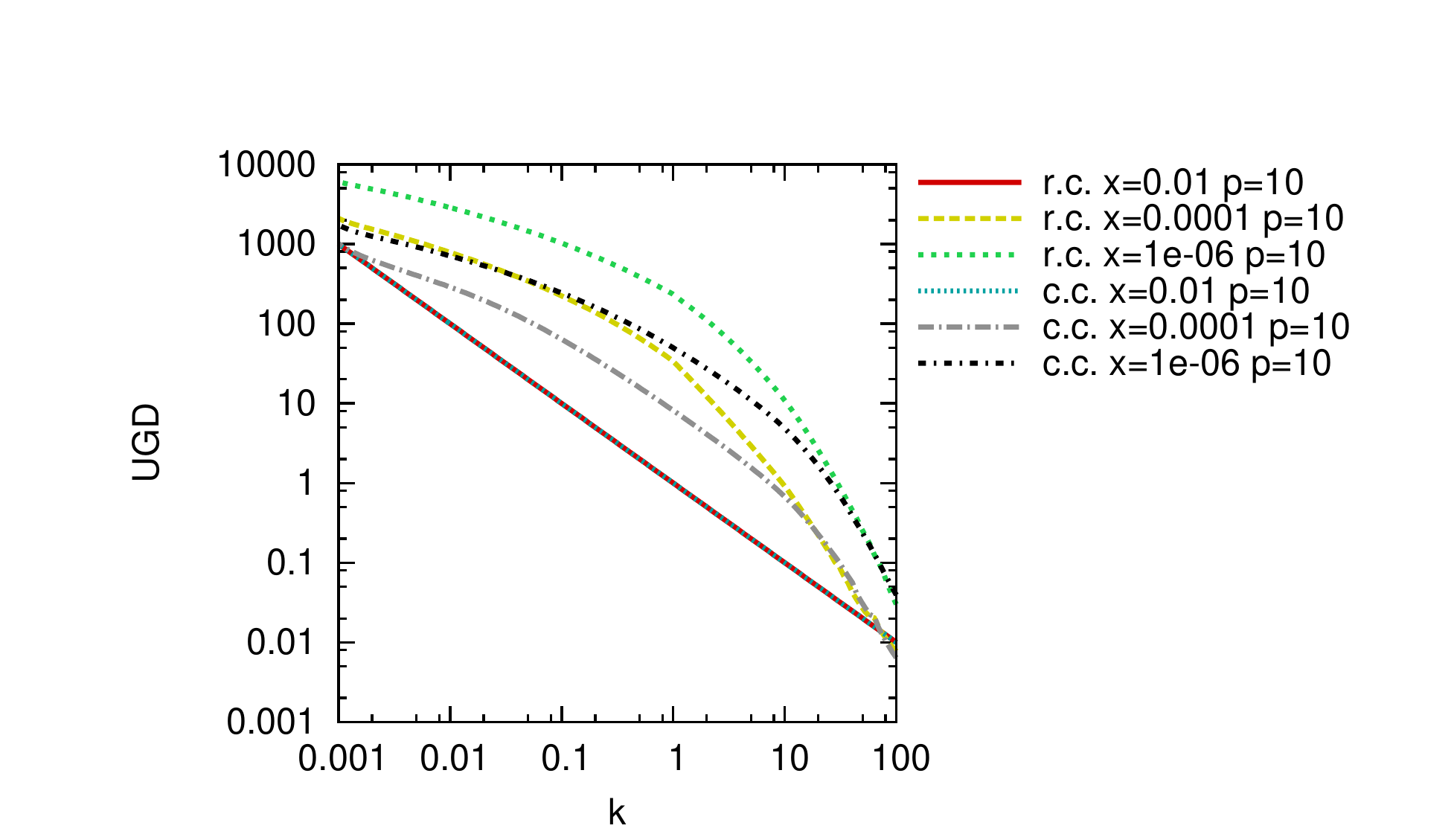}
}
\caption{Comparison of solutions of the KGBJS equation with constant ($\bar\alpha_s = 0.2$) and running coupling (Eq. \ref{running_alpha}).}
\label{kgbjs-const-vs-running}
\end{figure}
For the future phenomenological applications we investigate the effect of running coupling constant on the solution of considered equation.
The running coupling corrections were included in the following manner\footnote{In the future we are going to implement the running coupling constant as obtained in \cite{Kovchegov:2006wf}. For our present study this is however not crucial since our main point is the behavior of the saturation scale as a function of a hard scale and this should not change dramatically with another prescription for running of the coupling constant.}:
\begin{equation}\label{running_alpha}
\alpha_s(k^2)=
\frac{12\pi}{33 - 2 n_f}
\frac 1
{
  \ln
  {\frac
    {\max \left\{k^2, k_{freeze}^2\right\}}
    {\Lambda_{QCD}^2}
  }
}
\end{equation}
with $n_f=3$, $\Lambda_{QCD} = 0.2 \text{ GeV}$ and the $\max\left\{\cdot\right\}$ notation makes $k$ bounded by $k_{freeze} = 1 \text{ GeV}$.
The initial condition we choose to be:
\begin{equation}
\mathcal{E}_0(x, k, p) =
\frac
{\text{GeV}}
{k}
{\mathrm e}^{\bar\alpha_s(k^2)\ln\frac{x_0}{x}\ln\frac{k^2}{\mu^2}}.
\end{equation}
The extra $x$-dependent term is motivated by the resummation procedure for the BK equation and we use it also for the KGBJS equation in order to study differences in the evolution between these two equations. Its role is to attenuate the gluon density with decreasing $x$.
As we see on Fig. (\ref{kgbjs-const-vs-running}) showing the $x$ dependence of solutions at small $p$ considered form of the initial condition leads to falling distribution of the CCFM and KGBJS equations. This is not the case for the BK equation
as we see on Fig. (\ref{kgbjs-vs-bk}).
We see also on Fig. (\ref{kgbjs-const-vs-running}) that the effect of running coupling constant as compared to the fixed value at $\alpha_s=0.2$ leads to faster evolution and is more pronounced when the hard scale is larger.
The particularly interesting is the behavior of CCFM and KGBJS as a function of hard scale related variable $p$. The Fig. (\ref{hard-scale}) shows that the solution of the equations is a constant function of the $p$ variable as it is larger than transversal momentum of gluon. This effect can be understood by investigating the $\theta(p-z\,q)$ function in the considered equations. If the variable $p$ is larger than $k$ than the theta function sets to one and the angular ordering is relaxed. We expect this will have interesting implications for the saturation scale generated by the KGBJS equation.
\begin{figure}[t!]
\centerline{
\includegraphics[width=6.5cm,trim=0.75cm 0 4cm 0]
 {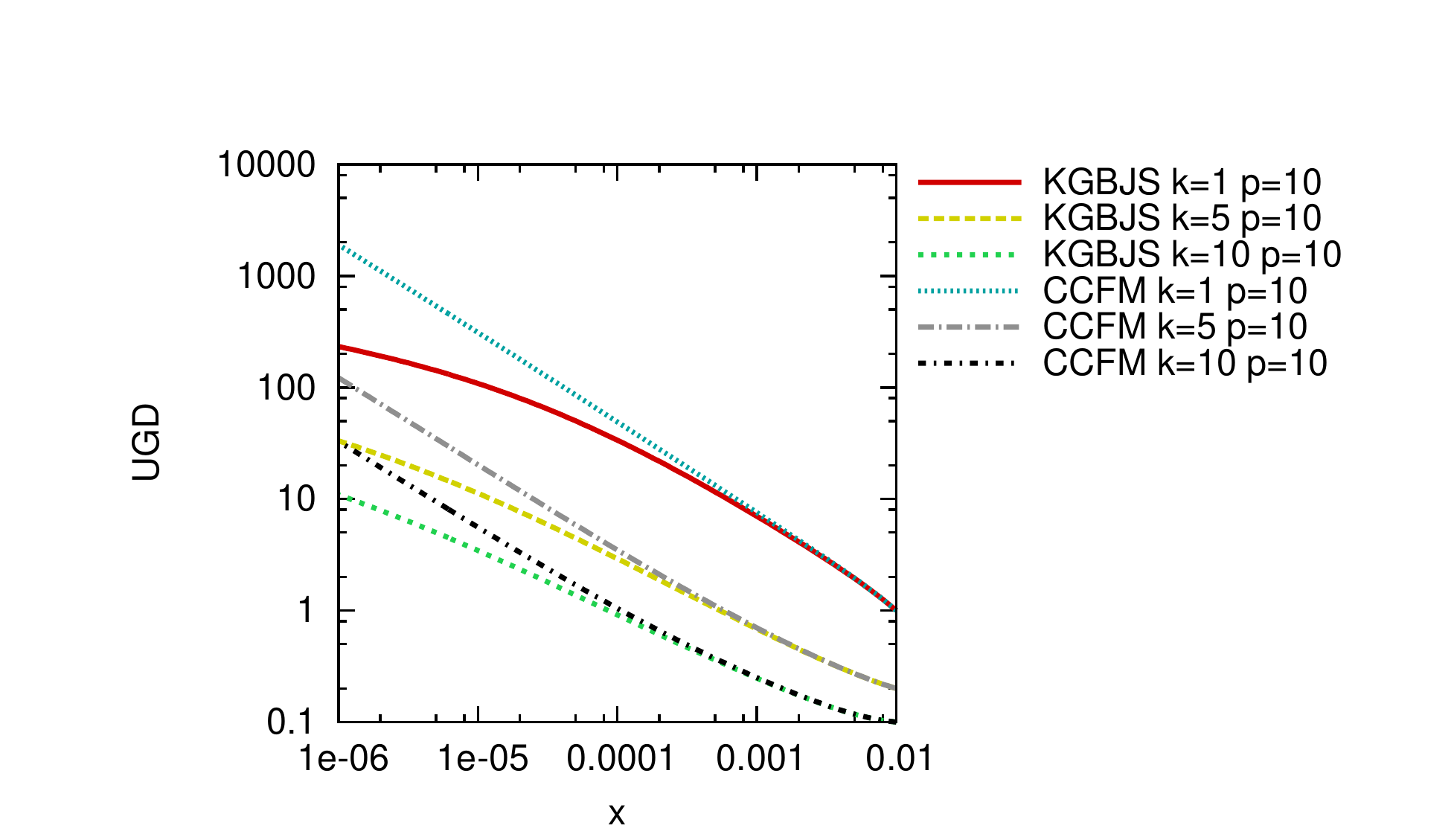}
 \includegraphics[width=6.5cm,trim=0.75cm 0 4cm 0]
 {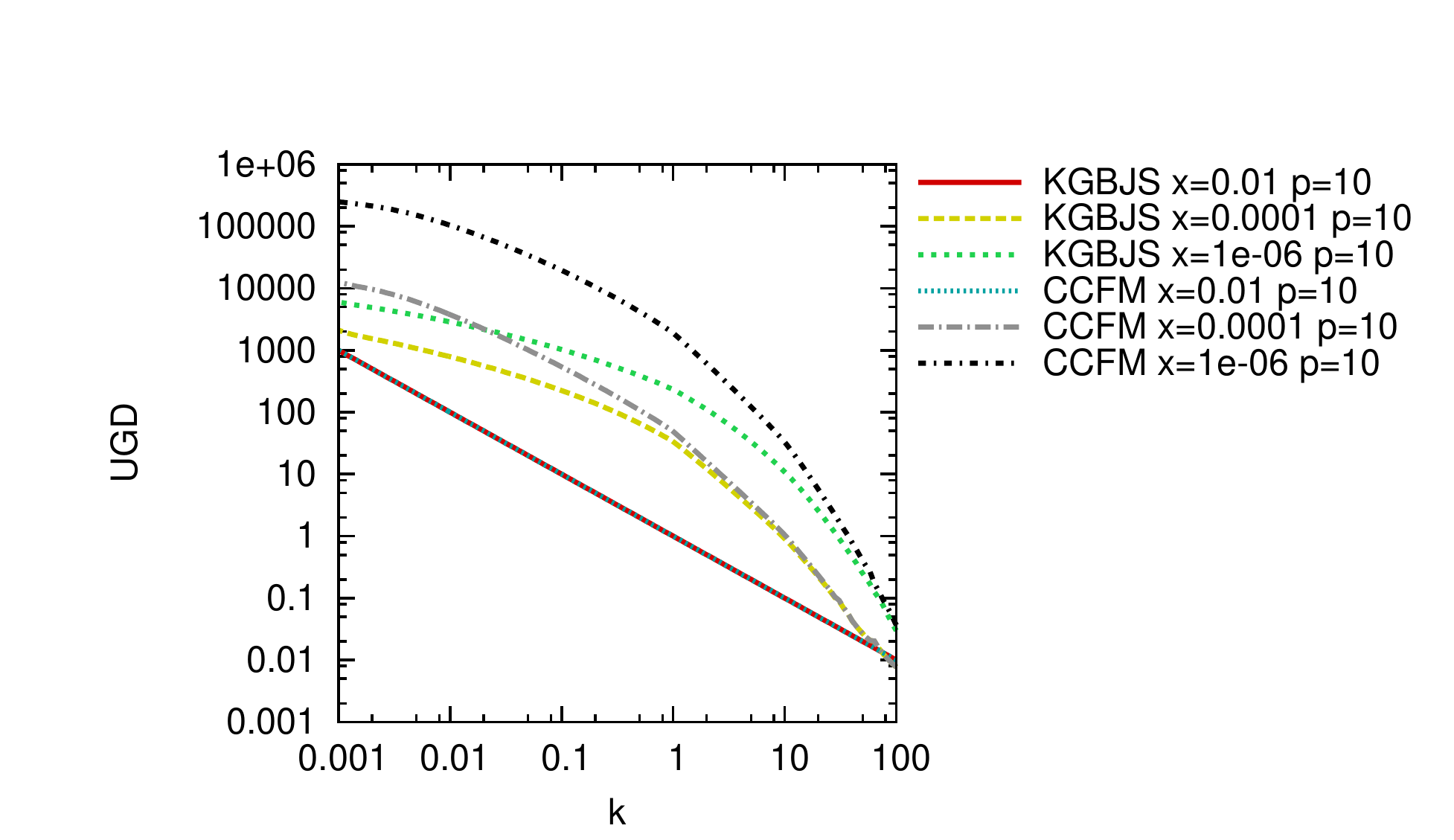}
}
\caption{Comparison of solutions of the KGBJS and CCFM equations (running $\bar\alpha_s$).}
\label{kgbjs-vs-ccfm}
\end{figure}
The plots on Fig.~(\ref{kgbjs-vs-ccfm}) compare solutions of CCFM and KGBJS.
We see the damping of the gluon density due to nonlinearity in case of KGBJS equation as we go towards low $x$ and low $k$ values.
\begin{figure}
\centerline{
 \includegraphics[width=6.5cm,trim=0.75cm 0 4cm 0]
 {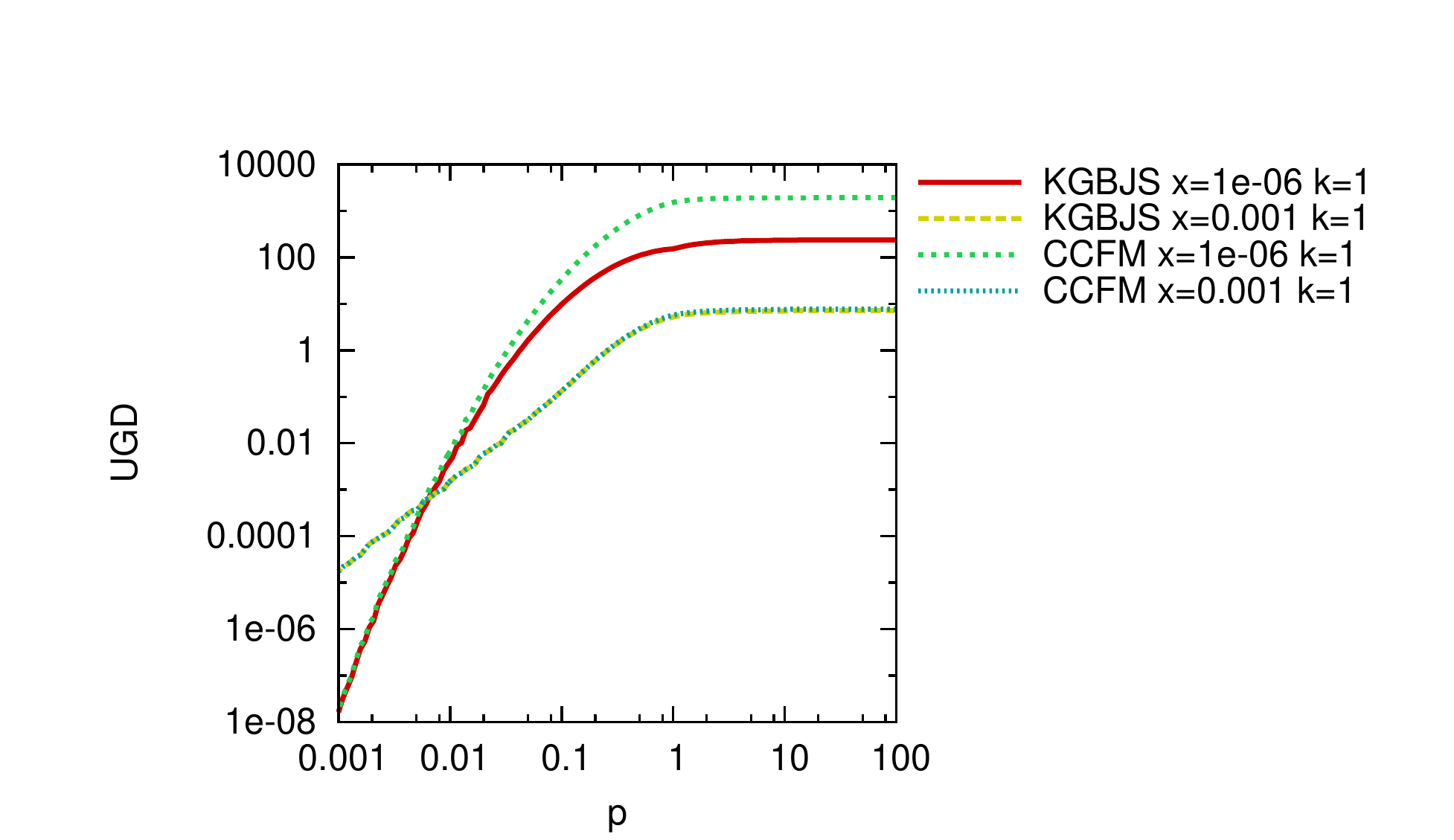}
}
\caption{Hard scale dependence of the CCFM and KGBJS equations.}
\label{hard-scale}
\end{figure}

\subsection{KGBJS and BK equations -- comparison}

\begin{figure}[t!]
\centerline{
 \includegraphics[width=6.5cm,trim=0.75cm 0 4cm 0]
 {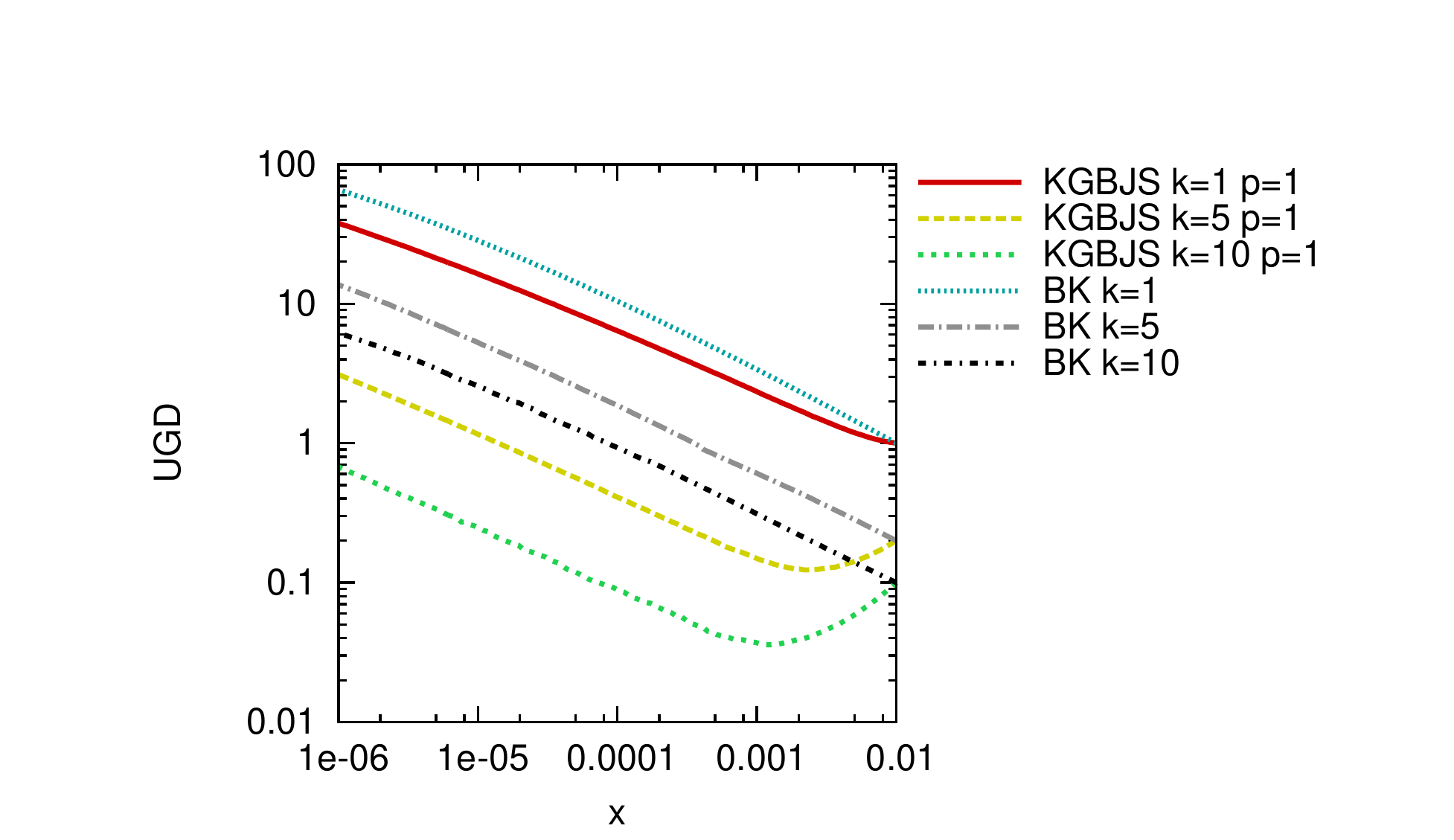}
 \includegraphics[width=6.5cm,trim=0.75cm 0 4cm 0]
 {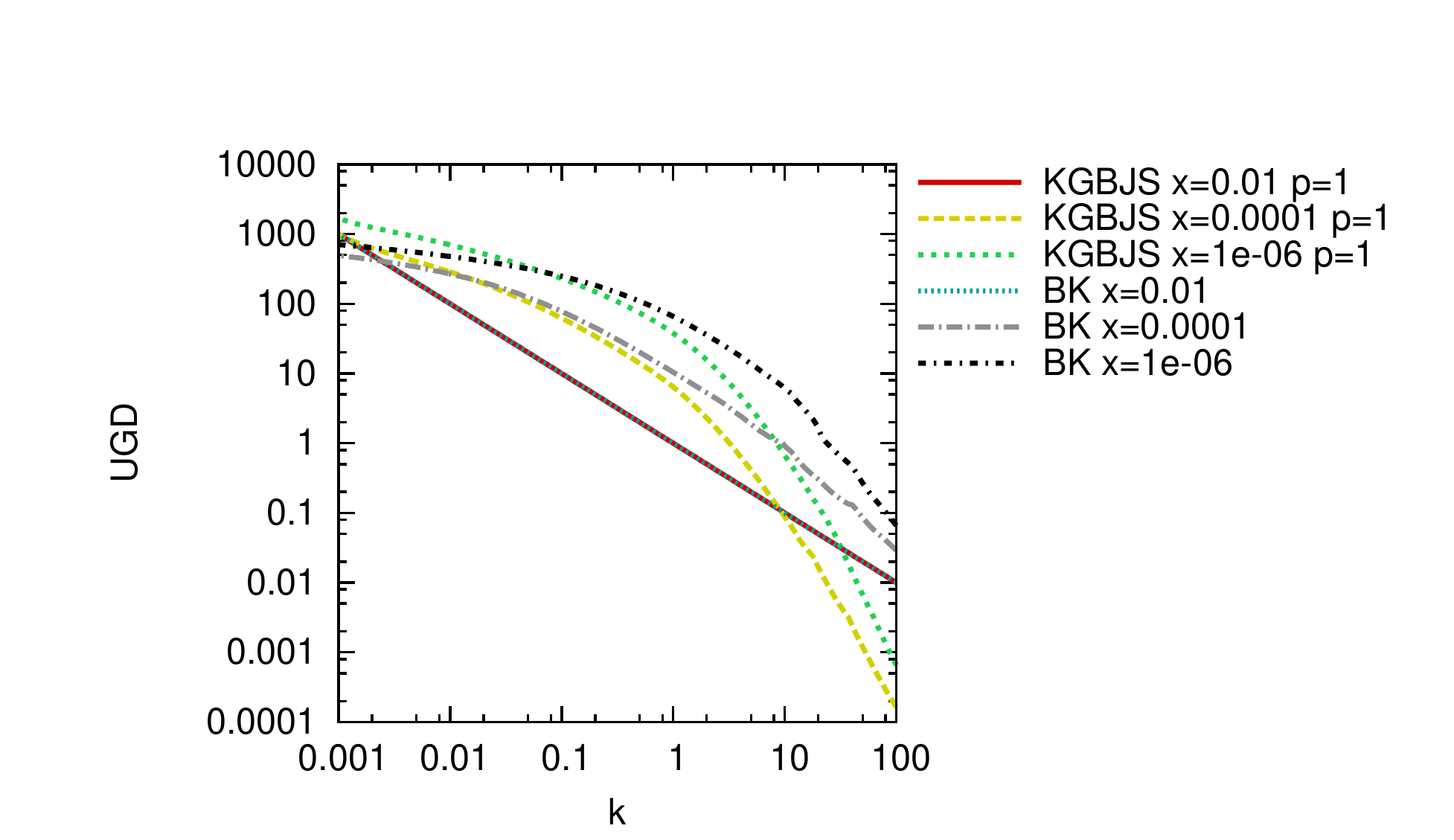}
}
\caption{Comparison of solutions of the KGBJS and BK equations (constant $\bar\alpha_s$).}
\label{kgbjs-vs-bk}
\end{figure}

The Balitsky-Kovchegov equation in the resummed form reads:
\begin{align}\label{eq:resBK}
\Phi(x,k^2)
&=\tilde \Phi^0(x,k^2)\\\nonumber
&+\overline\alpha_s
\int_{\frac{x}{x_0}}^1d\,z
\int\frac{d^2{\bf q}}{\pi q^2}\,
\theta(q^2-\mu^2)\frac{\Delta_R(z,k,\mu)}{z}
\Bigg[
\Phi(\frac{x}{z},|{\bf k} +{\bf q}|^2)\nonumber
\\\nonumber
&-\frac{1}{\pi R^2}q^2\delta(q^2-k^2)\,\Phi^2(\frac{x}{z},q^2)
\Bigg].
\end{align}
The Regge form factor assumes the form:
\begin{equation}
\Delta_R=e^{-\bar\alpha_s\ln1/z\ln k^2/\mu^2}
\label{eq:regge1}
\end{equation}
where $\mu$ is the resolution parameter. We assumed $\mu=0.01 \text{ GeV}$ in the calculations.
The equation above has been solved in \cite{Kutak:2013kfa} and its solution has been shown to be the same as the unresummed BK equation. It is instructive to compare the numerical solutions of the two equations in order to quantify the role of the angular ordering and dependence on the hard scale of the gluon density.
In Fig. (\ref{kgbjs-vs-bk}) we compare solutions of the KGBJS and the BK equations for fixed values of the coupling constant.
We see that as expected the slope of the solution of KGBJS equation (see Fig. \ref{kgbjs-const-vs-running}, right)
 is steeper due to suppression by the non-Sudakov form factor of large $k$ values. We also see that at low $k$ the saturation is weaker in the KGBJS equation as compared to BK. This could be understood by inspecting the nonlinear term of (\ref{eq:kgbjs2}). We see that if we perform the integration in the nonlinear part we obtain for the non-Sudakov form factor:
\begin{equation}
\Delta_{ns}(z,k,k)=e^{-\bar{\alpha}_s\ln^21/z}.
\end{equation}
This is to be compared with the Regge form factor in Eq. (\ref{eq:regge1}).
We see that for the fixed value of $k$ the nonlinear term in the KGBJS equation is more suppressed as compared to the BK equation therefore it leads to weaker saturation.
\section{Saturation of the exclusive gluon distribution}
To shed light on the importance of nonlinear corrections in the KGBJS,
we consider contour lines of the relative difference between solutions:
\begin{equation}
\beta(x, k, p) =
\frac
{\abs{\mathcal{E}_{CCFM}(x, k, p) - \mathcal{E}_{KGBJS}(x, k, p)}}
{\mathcal{E}_{CCFM}(x, k, p)}.
\label{eq:reldiff1}
\end{equation}
The traditional saturation scale $Q_s$,
i.e. transversal momentum for which the effects of nonlinearity are noticeable,
we define as:
\begin{equation}
\beta(x, Q_s(x, p), p) = const.
\end{equation}
Such quantity has been already defined for the BK equation \cite{Kutak:2004ym}:
\begin{equation}
\beta(x, Q_s(x)) = const
\end{equation}
with
\begin{equation}
\beta(x, k) =
\frac
{\abs{\Phi_{BFKL}(x, k) - \Phi_{BK}(x, k)}}
{\Phi_{BFKL}(x, k)}
\label{eq:reldiff2}
\end{equation}
where $\Phi_{BK}(x,k^2)$ is a solution of (\ref{eq:resBK}).

\begin{figure}
\centerline{
 \includegraphics[width=6.5cm,trim=0.75cm 0 4cm 0]
 {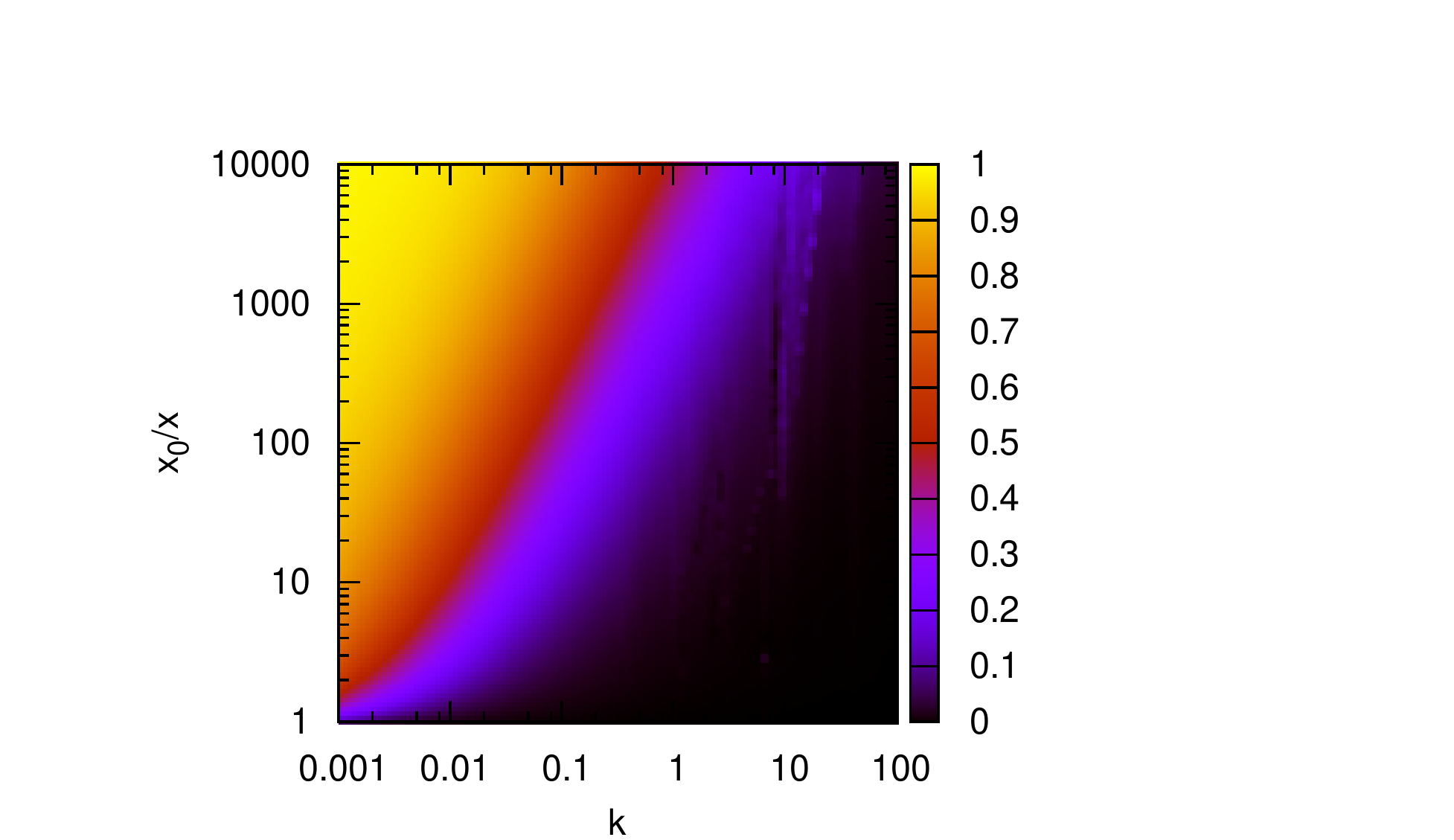}
}
\caption{Relative difference $\beta$ between solutions of BK and BFKL.}
\label{bk-beta}
\end{figure}

The quantity defined above, as observed in \cite{Kutak:2004ym}, has somewhat different slope compared to the saturation scale defined as a scale where the dipole amplitude is $1/2$.
However, as we see from the plots it is a good measure of the strength of nonlinearities.
The plot of $\beta$ on Fig. (\ref{bk-beta})
confirms the familiar growth of the saturation scale, which can be seen as $1/x$ is increasing upwards on the plot.

\begin{figure}[t!]
\centerline{
 \includegraphics[width=6.5cm,trim=0.75cm 0 4cm 0]
 {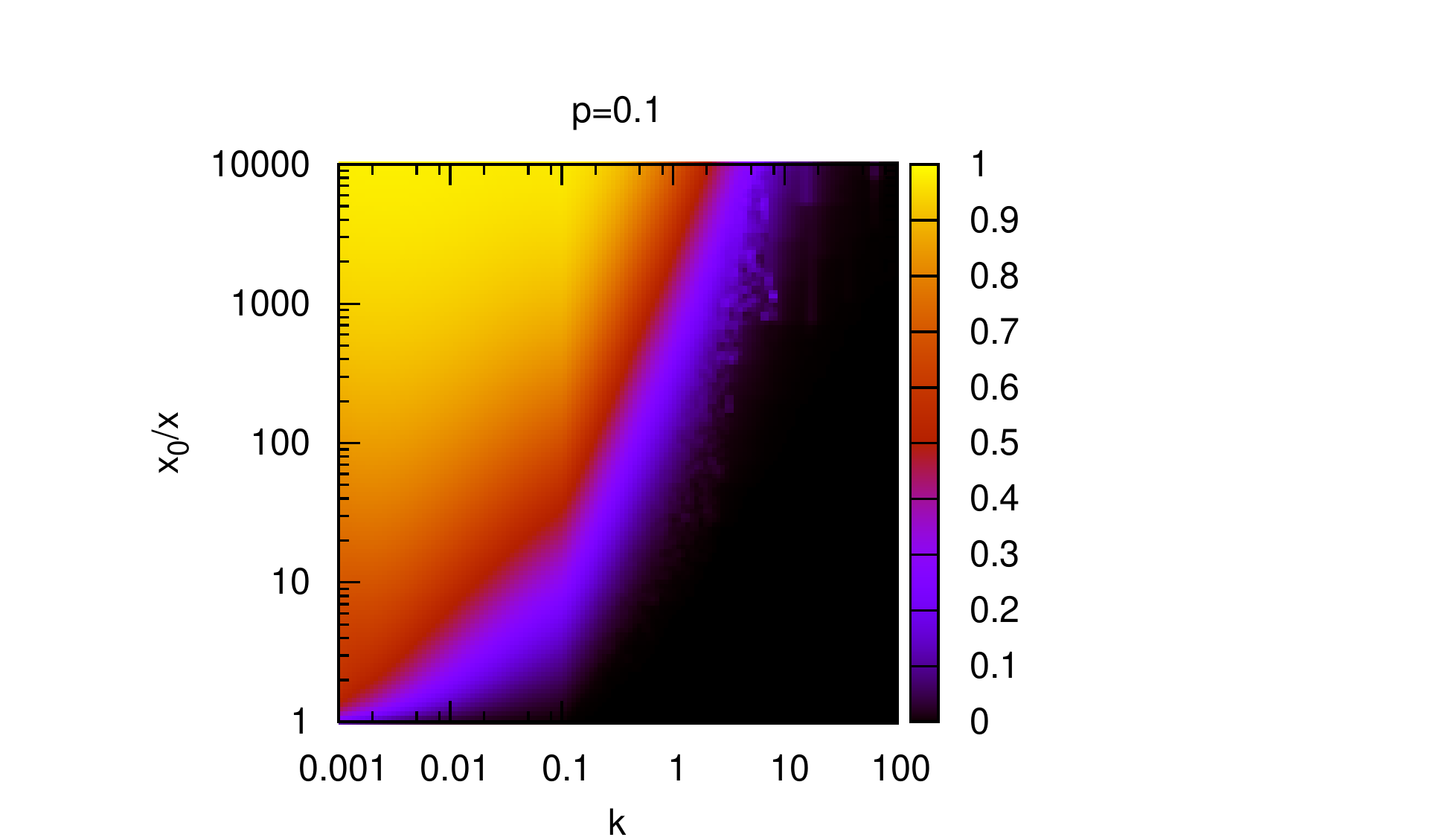}
 \includegraphics[width=6.5cm,trim=0.75cm 0 4cm 0]
 {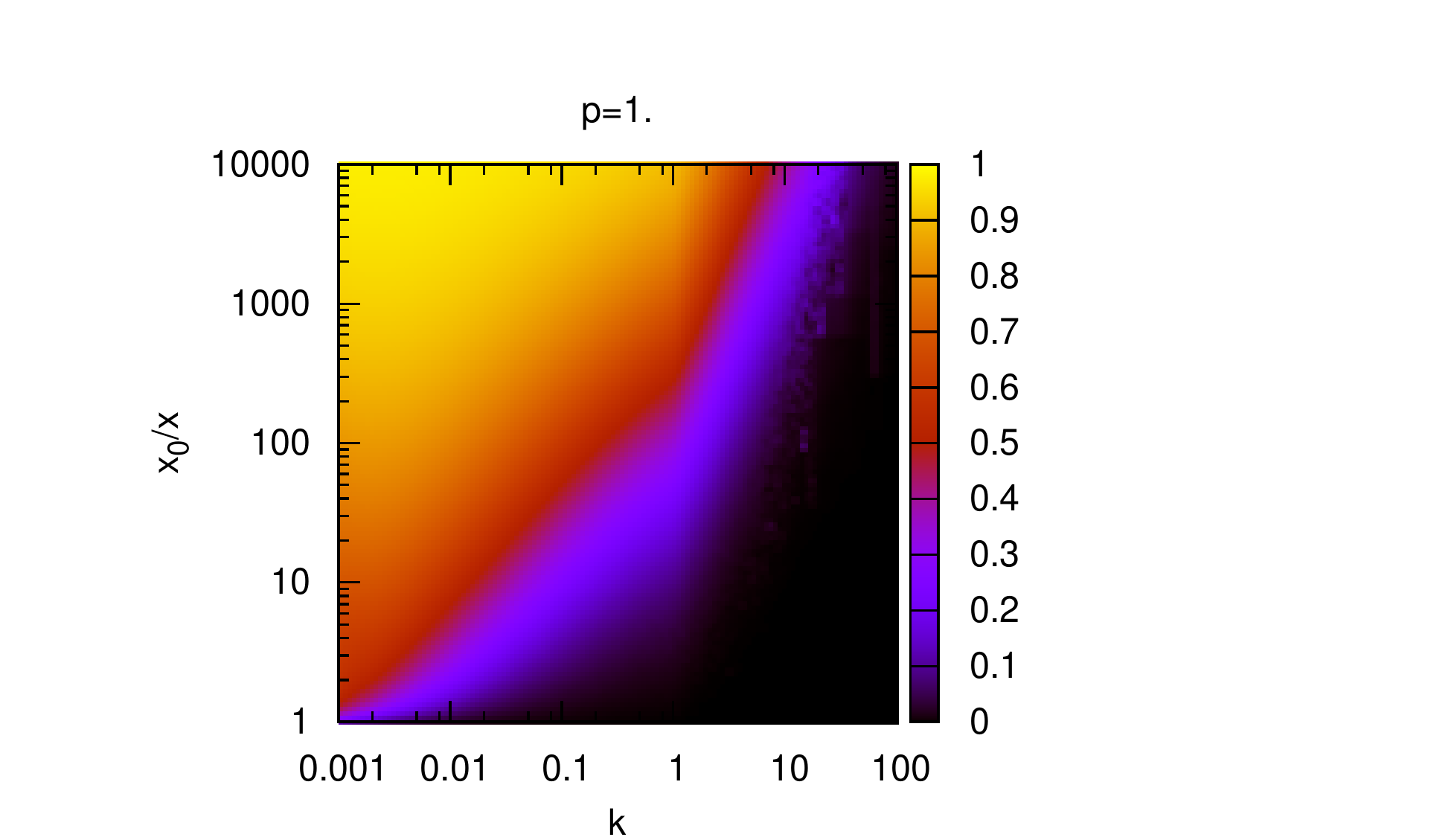}
}\centerline{
 \includegraphics[width=6.5cm,trim=0.75cm 0 4cm 0]
 {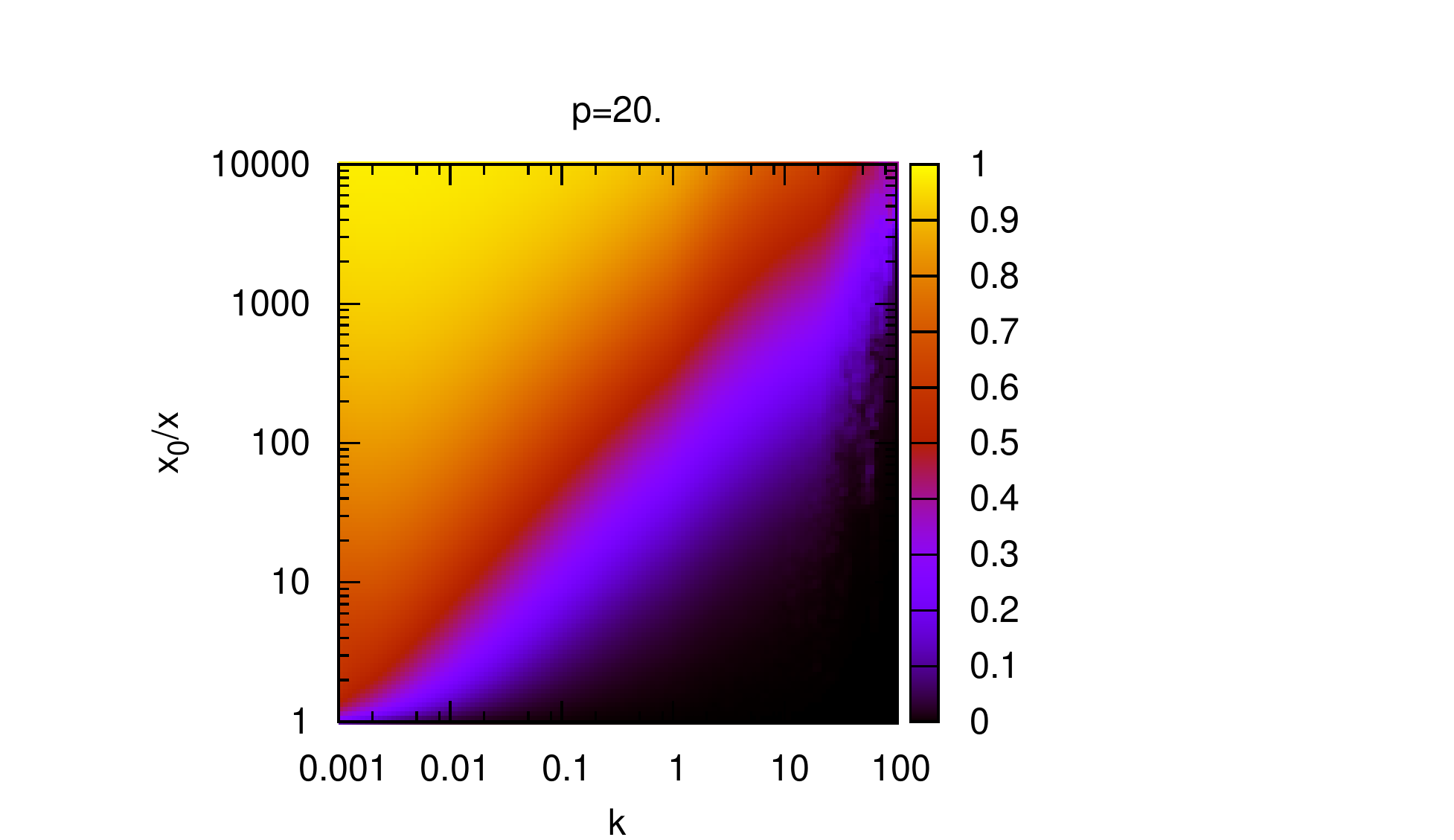}
 \includegraphics[width=6.5cm,trim=0.75cm 0 4cm 0]
 {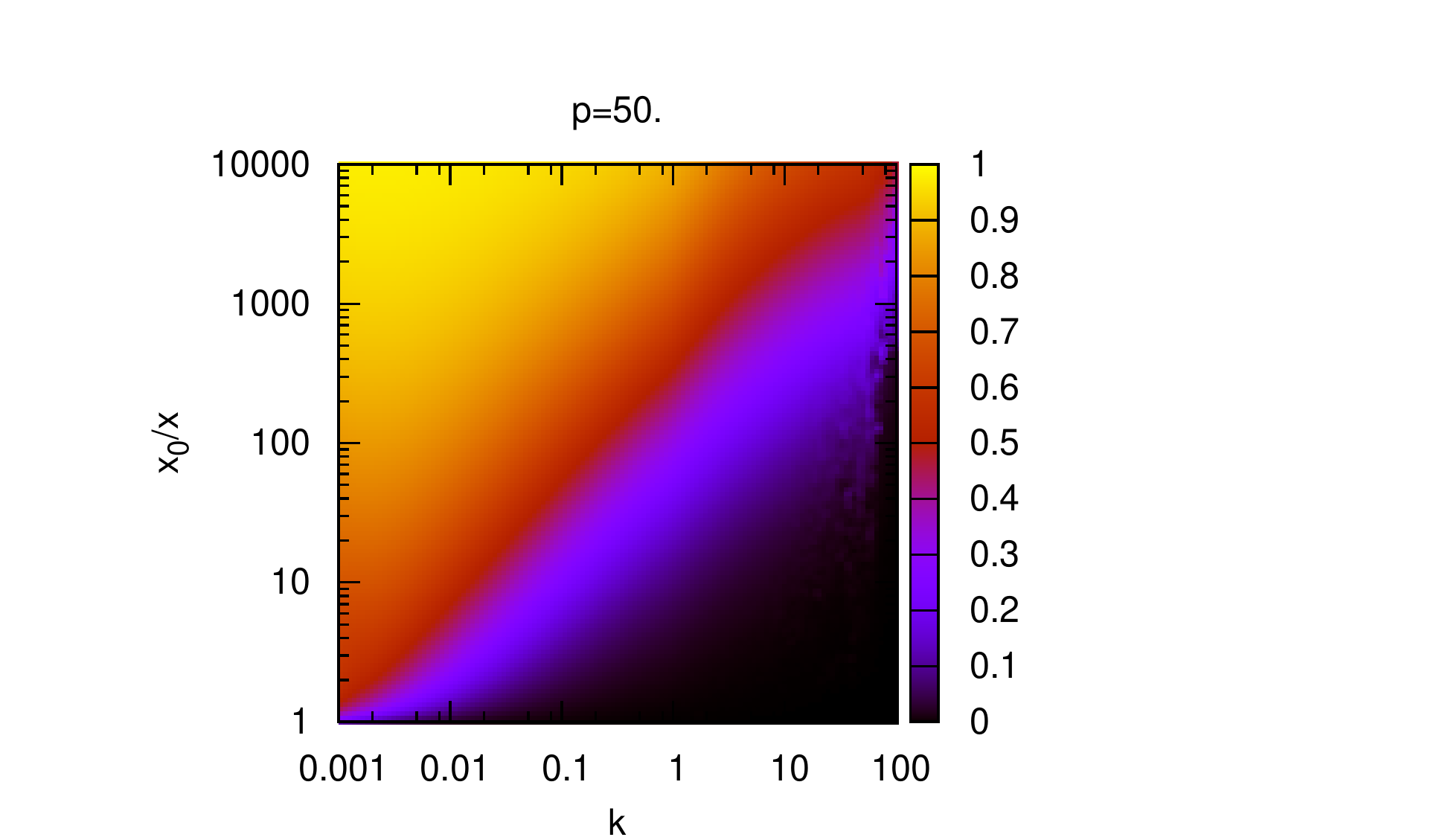}
}\centerline{
 \includegraphics[width=6.5cm,trim=0.75cm 0 4cm 0]
 {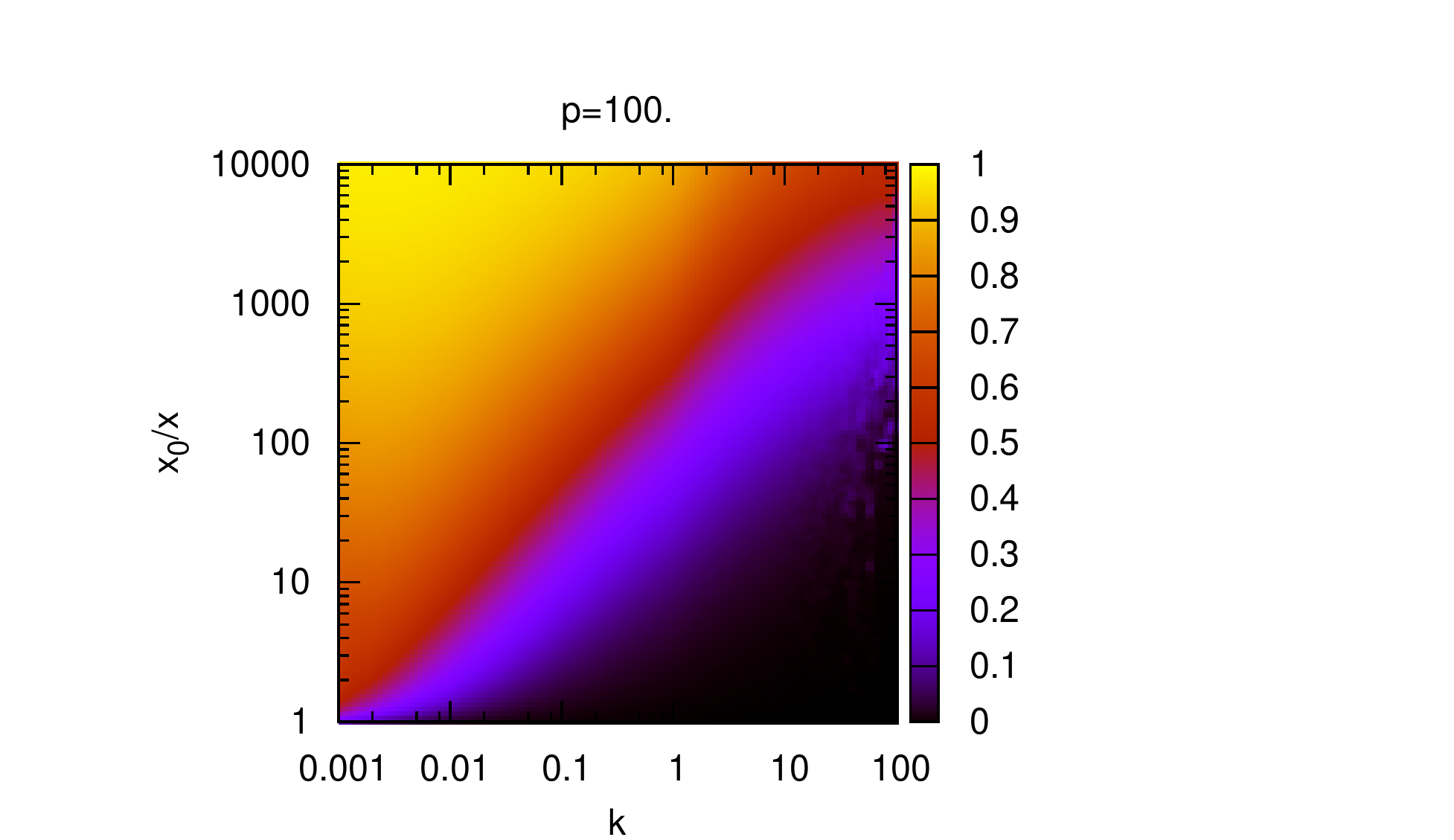}
}
\caption{The $\beta$ function (cross-sections for constant $p$). Solutions with running $\alpha_s$.}
\label{beta-alpharunning-p}
\end{figure}

\begin{figure}[t!]
\centerline{
 \includegraphics[width=6.5cm,trim=0.75cm 0 4cm 0]
 {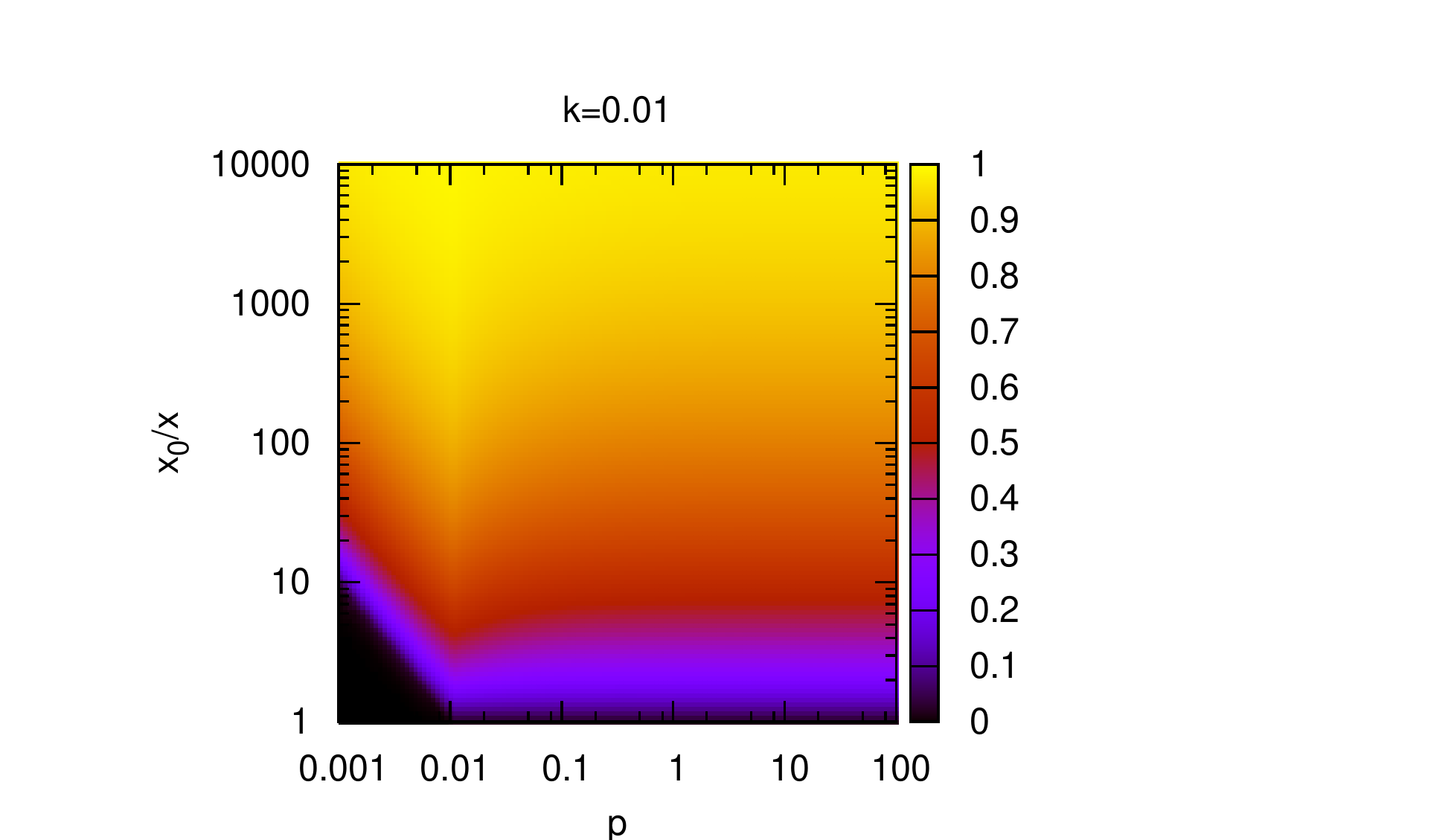}
 \includegraphics[width=6.5cm,trim=0.75cm 0 4cm 0]
 {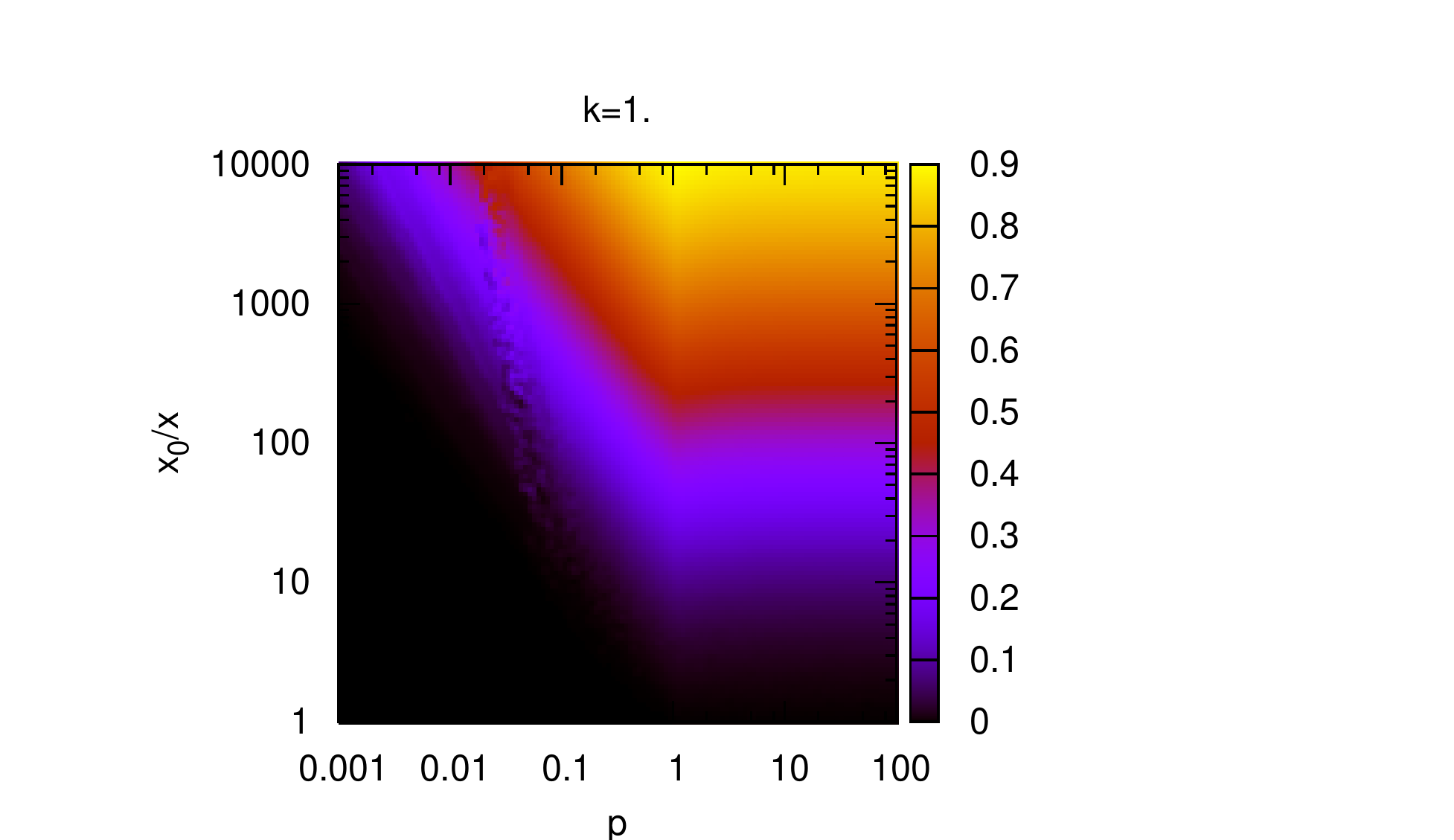}
}\centerline{
 \includegraphics[width=6.5cm,trim=0.75cm 0 4cm 0]
 {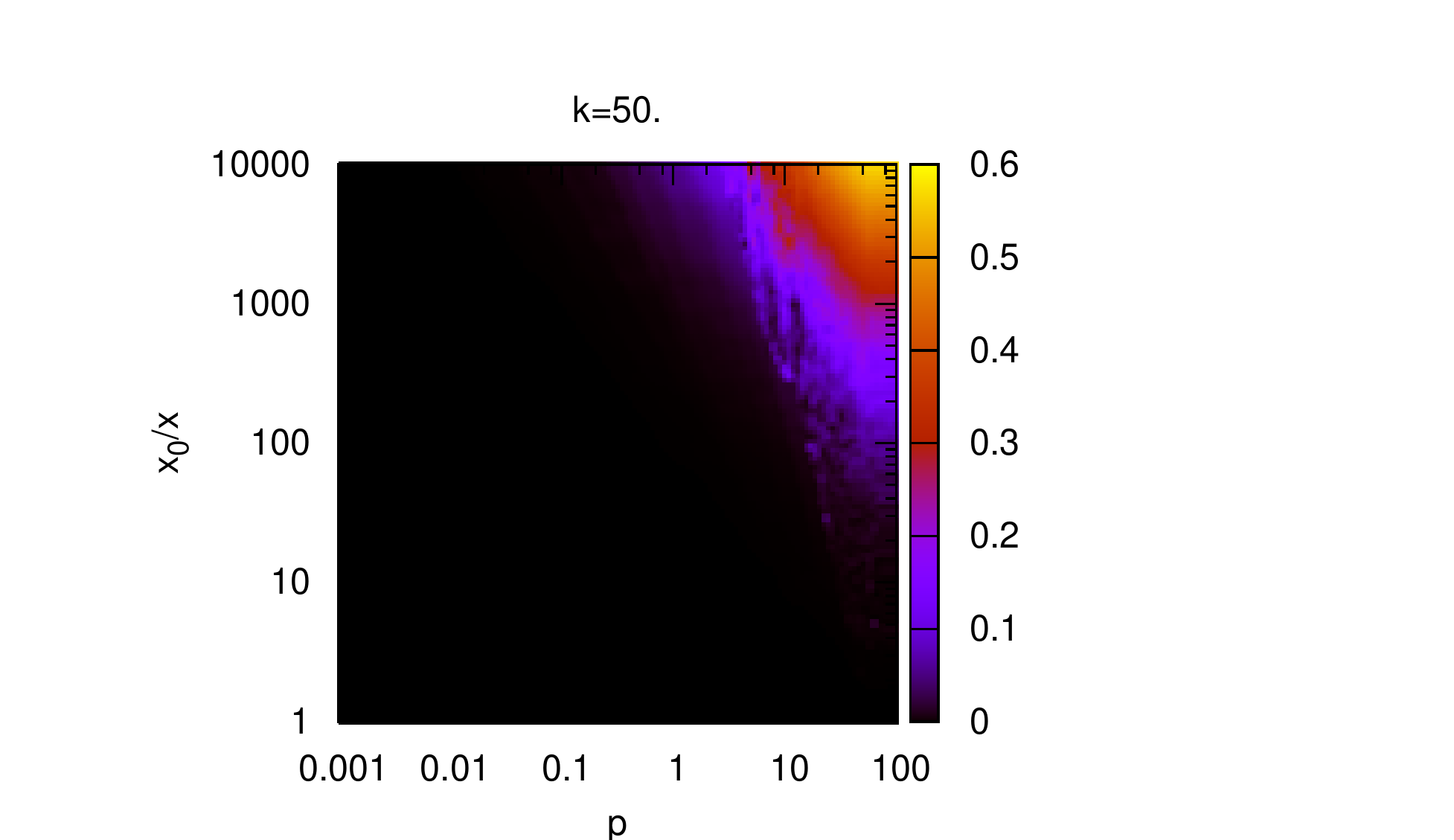}
}
\caption{The $\beta$ function (cross-sections for constant $k$). Solutions with running $\alpha_s$.}
\label{beta-alpharunning-k}
\end{figure}
The most interesting and novel effect as compared to the BK equation is the dependence of the saturation scale on the hard scale related variable $p$.
Several cross-sections of the $\beta$ function
(we limit ourselves to the running coupling case since the fixed coupling case does not bring anything new) on Figs. (\ref{beta-alpharunning-p}, \ref{beta-alpharunning-k})
indicate regions where KGBJS solutions diverge from results of the linear evolution.
The $k > p$ areas of the plots show
that the nonlinear effects enter when the $x_0/x$ is rather small. We also see that at $p\approx k$ the saturation line changes slope to larger value and as we go towards larger $k$ the saturation is weaker. However with growing $p$ the nonlinear effects become larger the slope becomes approximately constant and gluons get blocked by saturation.
This is the consequence of larger available phase space
(note the $\theta(p - z \bar q)$ factor in the kernel of the Eq. \ref{eq:final1})
for larger $p$ which allows for the gluon density to grow
and therefore to come at values where the nonlinear effects start to be important.
Eventually in phase space region where $p\gg k$ the KGBJS equation becomes independent on the hard scale and therefore the saturation scale stops to depend on it and gets liberated.  In this limit the maximal value of it is given and BK regime is reached. Similar effect has been already observed in \cite{Avsar:2010ia} with application of the absorptive boundary method (see for example Fig. (20) of \cite{Avsar:2010ia}, adjust it to have $Y$ axis vertical and compare it to presented here Fig. (\ref{beta-alpharunning-k})).
The difference is however in the strength of the effect since in the absorptive boundary method the authors of $\cite{Avsar:2010ia}$ set arbitrarily the value of gluon density below the saturation scale to a constant value while in our approach we allow for dynamical evolution and growth of gluon density.
The effect, called here liberation of saturation scale,
is linked to the so-called saturation of saturation scale expected in \cite{Kutak:2011fu, Avsar:2010ia}.
Since as we go towards the smaller values of $p$ we see that the saturation bends towards the $x_0/x$ axis and its growth is hindered.
Another aspect of the equation (\ref{eq:kgbjs2}) is that it allows
to define $p$-related saturation scale $P_s$ as:
\begin{equation}
\beta\left(x,k,P_s(x, k)\right) = const
\label{eq:pline}
\end{equation}
For fixed $k$, this function becomes a line, $P_s(x)$.
It indicates how the hard scale required to enter the saturation regime changes with $x$.
On Fig. (\ref{beta-alpharunning-k}) we plot the relative difference as defined in Eq. (\ref{eq:pline}) as a function of $p$ for varying $k$.
First of all we notice nontrivial relation between the values of $k$ and $p$ and nonlinearities.
If the value of $p$ is smaller than $k$
there is not much phase space available for growth of the gluon and the smaller $p$ is the lower $x$ has to be in order for the nonlinear effects to be visible. For values of $p>k$ the slope of the saturation region is roughly zero and for all values of $p$ the saturation enters at the same values of $x$.
We can say that the hard scale required to unveil the saturation
abruptly descends from infinity for some $k$-dependent $x$.
\section{Conclusions}
In this paper we performed numerical study of the simplified form of the KGBJS and CCFM evolution equations with running and fixed coupling constant. We compared the obtained solutions to the solution of the BK equation to investigate the interplay of saturation
and coherence.
We investigated the role of nonlinearity in the KGBJS equation
by studying the emergent saturation scale
i.e. the relative differences between solutions of the KGBJS and CCFM equations.
Due to the dependence of the KGBJS equation on the hard scale the saturation scale has been shown to depend on it in a nontrivial way.
In particular,
when the hard scale gets much larger than the $k$ of the gluon,
the saturation scale stops to depend on hard scale value
and liberates itself and is independent function of hard scale.
Finally we introduced hard scale related saturation scale $P_s$ i.e. measure of importance of nonlinearity as a function of hard scale and energy for fixed values of $k$. The analysis of the new scale shows that if the region when the $k$ of the gluon is larger than the hard scale the phase space is limited and the gluon density in order to be sensitive to nonlinear effects has to be evaluated at quite low $x$.
On the contrary if the scale $p$ is larger than gluon transversal momentum $k$
the $x$ values when the nonlinearities are important become quite large.
The presented analysis of the KGBJS equation is going to be extended in the future.
In particular, the impact on saturation of the Sudakov form factor and full splitting function is going to be investigated as well as the properties of solution of equation written directly for the unintegrated gluon density in \cite{Kutak:2012qk}.
\section*{Acknowledgments}
We would like to thank K. Bozek, M. Deak, K. Golec-Biernat, H. Jung, R. Peschanski, W. Placzek, K. Slawinska for useful discussions.\\
Krzysztof Kutak and Dawid Toton were supported during this research by  Narodowe Centrum Badań i rozwoju with grant LIDER/02/35/L-2/10/NCBiR/2011.

\end{document}

%% file: common_preamble.tex
\usepackage[utf8x]{inputenc}
\usepackage[T1]{fontenc}
\usepackage{dsfont}
\usepackage{amstext}
\usepackage{amsfonts}
\usepackage{amsmath}
\usepackage{graphicx}
\usepackage[extendedchars]{grffile}
\usepackage{setspace}
\usepackage{braket}
\usepackage{gensymb}

\newcommand{\abs}[1]{\left|#1\right|}

\newcommand{\dd}{\mathrm d}